\RequirePackage[hyphens]{url}
\documentclass[journal=ancac3,manuscript=article]{achemso}
\usepackage[version=3]{mhchem} 
\usepackage{xcolor}
\usepackage{soul} 
\usepackage{amssymb}
\usepackage{lineno,hyperref}
\usepackage{graphicx}
\usepackage{bm}
\usepackage{float}

\newcommand{\DIPC}[0]{
Donostia International Physics Center (DIPC),
Paseo Manuel de Lardizabal 4, 20018 Donostia-San Sebasti\'an, Spain}
\newcommand{\CFM}[0]{
Centro de F\'{\i}sica de Materiales CFM/MPC (CSIC-UPV/EHU), Paseo Manuel de Lardizabal 5, 20018 Donostia-San Sebasti\'an, Spain}
\newcommand{\PolymerEHU}[0]{Departamento de Pol\'{i}meros y Materiales Avanzados: F\'{i}sica, Qu\'{i}mica y Tecnolog\'{i}a, Facultad de Qu\'{i}mica (UPV/EHU), Apartado 1072, 20080 Donostia-San Sebasti\'{a}n, Spain}
\author{Natalia E. Koval}
\email{natalia.koval@ehu.eus}
\affiliation{\CFM}

\author{J. I\~{n}aki Juaristi}
\affiliation{\PolymerEHU}
\alsoaffiliation{\CFM}
\alsoaffiliation{\DIPC}

\author{Maite Alducin}
\affiliation{\CFM}
\alsoaffiliation{\DIPC}
\title{Strong-field effects in the photo-induced dissociation of the hydrogen molecule on a silver nanoshell}

\date{\today}

\begin{document}



\begin{abstract}Plasmonic catalysis is a rapidly growing field of research, both from experimental and computational perspectives. Experimental observations demonstrate an enhanced dissociation rate for molecules in the presence of plasmonic nanoparticles under low-intensity visible light. The hot-carrier transfer from the nanoparticle to the molecule is often claimed as the mechanism for dissociation. However, the charge transfer time scale is on the order of few femtoseconds and cannot be resolved experimentally. In this situation, \textit{ab initio} non-adiabatic calculations can provide a solution. Such simulations, however, have their own limitations related to the computational cost. To accelerate plasmonic catalysis simulations, many researchers resort to applying high-intensity external fields to nanoparticle-molecule systems. Here, we show why such an approach can be problematic and emphasize the importance of considering strong-field effects when interpreting the results of time-dependent density functional theory simulations of plasmonic catalysis. By studying the hydrogen molecule dissociation on the surface of a silver nanoshell and analyzing the electron transfer at different field frequencies and high intensities, we demonstrate that the molecule dissociates due to multiphoton absorption and subsequent ionization.\end{abstract}



\section{Introduction}

Plasmon-induced photocatalysis (plasmon-enhanced nanocatalysis or plasmonic catalysis) has emerged as a highly promising field that combines the unique properties of plasmonic nanoparticles (NPs) with catalytic processes.
\cite{Linic2011,Linic2015,acswc,Zhang_2013,Wei2018,Zhang2019,Cortes2020,developmentsinplasmonassistedphotocatalysis2020,Kumar2021,Newmeyer2022,Jain2022,pubs.acs.org/acscatalysis2023,Dong2023,Amirjani2023} NPs have been widely explored as catalysts due to their large surface-to-volume ratio, tunable surface properties, and localized surface plasmon resonance (LSPR) effects.
These properties enable the manipulation of light-matter interaction and the generation of highly localized electromagnetic fields (hotspots) and energetic charge carriers (hot electrons), leading to enhanced catalytic activity. \cite{Zhang_2013,Amirjani2023,Li2023a,Brooks2018,LucasV.Besteiro2023,ConstantinosMoularas2024} 
The choice of NP material, shape, and size plays a critical role in determining their catalytic performance. \cite{Huynh2022,Amirjani2023} 
Some common metals used for plasmonics include silver, gold, and copper. In addition, plasmonic NPs come in various shapes, most commonly being spherical NPs, \cite{Jin2022} but also nanorods and nanostars, \cite{SinhaRoy2017, Atta2019} nanotriangles, \cite{Thangamuthu2019} nanocubes, \cite{Huynh2022,Herring2023} and nanoshells. \cite{doi:10.1021/la3050626,www.acsami.org2023,Halas2005,Kang2014,Huang2016,Srinoi2020,LaurentLermusiaux2023}

Experimentally, plasmonic catalysis has been demonstrated by several groups. \cite{acswc,Christopher2011,Mukherjee2012,Brongersma2015, Boerigter2016a} Mukherjee et al. \cite{Mukherjee2012} studied plasmon-induced dissociation of H$_2$/D$_2$ on Au/TiO$_2$ at room temperature. Supporting their findings by DFT calculations, the authors suggested a dissociation mechanism consisting in hot-electron transfer from Au to the molecular antibonding state facilitated by H$_2$-Au hybridization.
Christopher et al. \cite{Christopher2011} reported
enhanced performance in the oxidation of ethylene on Ag nanocubes. They also suggested that hot electrons transfer to the lowest unoccupied molecular orbital (LUMO) and dissipate energy into the vibrational modes of the molecule, stretching the O$_2$ bonds and eventually activating dissociation. 

Employing advanced computational methods like real-time time-dependent density functional theory (RT-TDDFT), \cite{Runge1984,2006-book-tddft,ullrich2011time} makes it possible to unravel phenomena occurring in very short times, often difficult to resolve experimentally. RT-TDDFT is a powerful tool for modeling light-matter interaction, plasmonic properties, and catalytic processes. \cite{Herring2023a}
In this respect, there is an increasing number of studies using RT-TDDFT combined with nonadiabatic Ehrenfest molecular dynamics (EMD) that report plasmon-assisted dissociation of molecules near plasmonic NPs. \cite{Yan2015,YuZhang2018,Hull2020,YiminZhang2021,Huang2021,KudaSingappulige2023,Zhang2021b,Herring2023,Li2023c,Wang2023c} However, these kinds of studies are still scarce and very specific, making it difficult to extract more general conclusions regarding the mechanisms ruling the dissociation process. In particular, most of the publications only focus on resonant frequencies of the external field. \cite{YuZhang2018,Hull2020,Feng2022,Hull2023,YiminZhang2021,zhang22,Herring2023} There is a limited number of studies extending the analysis to non-resonant frequencies. A good example is the work by Yan et al.,~\cite{Yan2015} showing that H$_2$ adsorbed on one end of an Ag$_6$ chain dissociates when external field frequency $\omega_0$ coincides with LSPR ($\omega_\textrm{p}$), but not at lower ($\omega_0 < \omega_\mathrm{p}$) or higher ($\omega_0 > \omega_\mathrm{p}$) frequencies. It is, hence, essential to perform more studies comparing frequencies both in and out of resonance with $\omega_\textrm{p}$ to demonstrate the role of plasmon excitations in catalytic reactions.

The role of field intensity in the induced reactivity is another issue that requires further consideration. All the aforementioned RT-TDDFT-EMD research uses very high field intensities and report on a threshold intensity for dissociation to occur, which lies in the range of $10^{13}-10^{15}$~W/cm$^2$ (with the pulse duration 10-60~fs). \cite{Yan2015,YuZhang2018,Hull2020,Hull2023,YiminZhang2021,zhang22,Herring2023, Wang2023c} Applying strong fields may lead to nonlinear effects such as high-harmonic generation, \cite{Andreev2013} above-threshold ionization, multiphoton ionization, and tunneling. \cite{M_Protopapas_1997,CALVAYRAC2000493,Gruzdev2014,Konda2021,Alemayehu2023,Reutzel2020,Bionta2016,Saydanzad2022,Saydanzad2023Nanoph} 
Unfortunately, strong-field effects are typically not discussed in the RT-TDDFT-EMD literature. 
The latter may be due to technical limitations of the methodology to describe properly those additional effects.
For instance, using atom-centered basis sets, \cite{Hull2020,Hull2023,Herring2023} although is computationally efficient, does not allow for the account of ionization due to the absence of any basis functions to represent unbound electrons.
In contrast, the existing studies using real-space representation are able to describe ionization and emission into the continuum, but rarely discuss strong-field effects. \cite{Yan2015,YuZhang2018,YiminZhang2021,Huang2021,zhang22} Yan et al. \cite{Yan2015} briefly mention some electron loss without analyzing its effect on the dissociation process. Huang et al. \cite{Huang2021} obtain for H$_2$O on AuNPs that above $I_\mathrm{max} = 1.34\times 10^{14}$ W/cm$^2$, the linear dependence of the H$_2$O splitting on intensity breaks. The latter is attributed to nonlinear coupling of the external field to the system that may lead to multiphoton absorption and subsequent water fragmentation. At lower intensities, however, they suggest water splitting by hot-electron transfer from the NP to the molecular antibonding orbital. The maximum dissociation rate was observed not at the plasmon frequency, but at lower one corresponding to the energy gap between the Fermi level and the antibonding state.

The field of plasmon-induced photocatalysis is certainly very active and there are still many open questions to solve. Actually, the precise mechanism of the plasmon-activated dissociation of molecules has yet to be understood. \cite{acswc,Herring2023a} In this respect, more research is needed to get further insights into the limitations of {\it ab initio} simulations and the precise modeling and external field conditions. 
With this motivation in mind, we present here a systematic study of H$_2$ dissociation induced by a silver hollow nanoshell of the Ag$_{55}$ NP, (hereafter denoted as Ag$_{55}^{L1}$, with $L1$ standing for "layer 1", i.e., the outer layer of Ag$_{55}$). Silver NPs are known for their high plasmonic activity and strong LSPR in the visible region. \cite{Liang2020} For near-spherical AgNPs, the experimental plasmon frequency
varies from 3 to 4~eV ($\approx 400-300$~nm wavelength) depending on size, thus it lies mostly in the UV range.~\cite{Yu2020,Schira2019} Constructing nanoshells is another way of tuning the plasmon resonance. \cite{Halas2005} Linear-response TDDFT calculations have shown that the plasmon frequency of Ag hollow shells experiences a redshift compared to that of AgNPs of the same diameter.~\cite{Koval2016} 
Thus, it is a practical way of shifting $\omega_\textrm{p}$ into the visible range and reducing the computational cost because of the smaller number of atoms at the same diameter. We analyze in detail the underlying
mechanisms and conditions of molecular dissociation and highlight some important limitations of the currently accepted modeling approach. We discuss similarities of our results with the literature focusing on the implications of applying a strong field to plasmonic systems. We improve upon the limitations of RT-TDDFT with localized basis set and demonstrate that including strong-field effects changes our conclusions about the mechanism of dissociation. Namely, we show that the ionization of the molecule leads to its dissociation and desorption from the surface of the nanoshell at high field strength and frequency, regardless of the plasmon resonance obtained in linear regime. We emphasize that applying strong external fields to plasmonic nanoparticles leads to nonlinear effects that play the principal role in molecular dissociation, overshadowing the expected influence of plasmonic effects.

Furthermore, it is worth to mention that EMD, due to its mean-field nature, has inherent limitations, particularly when multiple pathways for nuclear dynamics are available.~\cite{doi:10.1021/cr3004899,10.1063/1.5055768} In such cases, the mean-field approximation can underestimate certain dissociation pathways by averaging over electronic states. Recent works have applied an alternative approach based on the surface-hopping method,~\cite{10.1063/1.459170,Wu2022,Wu2023} which allows for transitions between different potential energy surfaces and accounts for the branching of trajectories due to electron-nuclear coupling. However, these limitations of EMD are not critical for our specific study. Since our primary goal is to investigate the electronic response to strong external fields rather than to calculate dissociation probabilities, EMD combined with RT-TDDFT is expected to perform well. Moreover, the short timescale, system size, and strong-field conditions explored in this work justify the use of RT-TDDFT combined with EMD to capture the key nonlinear processes relevant to plasmonic catalysis.~\cite{annurev-physchem-040214-121359,Herring2023a}

\section{Methods}
	
\subsection{Geometry optimization}
	
The first step in our computational approach involves the optimization of the nanoshell geometry using density functional theory (DFT). We employed the Perdew-Burke-Ernzerhof (PBE) functional \cite{Perdew1996} within the CP2K software package, \cite{Hutter2013,Kuehne2020,cp2k} which implements the Gaussian plane wave (GPW) method. \cite{LIPPERT1997,VandeVondele2005} We used DZVP basis set including 11 electrons for Ag explicitly. A cutoff of 600 Ry was used for the grid. Norm-conserving Goedecker-Teter-Hutter (GTH) pseudopotentials \cite{Goedecker1996} were used to represent the interaction of valence electrons with atomic cores. The initial coordinates of the nanoshell were obtained using Atomic Simulation Environment (ASE) builder (function ase.cluster.Icosahedron) \cite{HjorthLarsen2017} and taking only the outer layer of the Ag$_{55}$ icosahedral cluster, which contains 42 atoms. We denote the nanoshell as Ag$_{55}^{L1}$. The geometry optimization was carried out until the maximum force on each atom was below 0.001 Hartree Bohr$^{-1}$. Additionally, geometry optimization was performed for Ag$_{55}^{L1}$+H$_2$ placing H$_2$ at a distance 2 {\AA} from the cluster facet along the $z-$axis. Final distance between the nanoshell facet and the molecule after the optimization is 3 {\AA}. A non-periodic simulation cell of $20\times20\times20$ \AA$^3$ was used in the simulations.
	
\subsection{Real-time time-dependent density functional theory calculations of the absorption spectrum}

The absorption spectrum was computed within the real-time time-dependent density functional theory (RT-TDDFT) approach implemented in the CP2K software package. \cite{Hutter2013,Kunert2003,Andermatt2016,Kuehne2020,cp2k,LIPPERT1997,VandeVondele2005} We employed the enforced time reversible symmetry (ETRS) real-time propagation scheme. Non-periodic boundary conditions were used with a unit cell of $20\times20\times20$ \AA$^3$. We applied a small perturbation (a $\delta$-kick of field strength of 0.001 a.u.) to the system along the $x$, $y$, and $z$ directions at $t = 0$. Next, we let the system evolve in time during 6000 simulation steps with a time step $\Delta t = 0.005$~fs. Absorption spectra in the frequency domain were computed by applying the discrete Fourier transform to the time-dependent dipole moment in each direction and then calculating the average spectrum over the three directions. The three components are almost identical due to the symmetry of the nanoshell.
	
We validated our computational approach by comparing the computed absorption spectrum for a full Ag$_{55}$ icosahedral cluster with available calculations from the literature. Our plasmon peak at 3.8 eV is in good agreement with other TDDFT calculations, \cite{Weissker2011,Rossi2015,Baseggio2016,Koval2016} indicating the reliability of the methodology employed here.
Comparing to the full cluster, the spectrum of the nanoshell experiences a red shift, in agreement to what has been observed in ref \citenum{Koval2016}.
	
\subsection{Ehrenfest molecular dynamics simulations}

To investigate the action of the external field on the Ag$_{55}^{L1}$+H$_2$ system, we performed RT-TDDFT simulations combined with Ehrenfest molecular dynamics (EMD) implemented in the CP2K software package.\cite{Hutter2013,Kunert2003,Andermatt2016,Kuehne2020,cp2k,LIPPERT1997,VandeVondele2005} A converged time step $\Delta t = 0.002$ fs was used in all the RT-TDDFT-EMD simulations. Each simulation was run for 55 fs. The external field was modeled by a Gaussian envelope (see Figure \ref{fgr:str_abs_spec}(c)):
\begin{equation}
    E(\omega_0,t) = E_0~ \mathrm{exp}\left[-\frac{(t-t_0)^2}{2\sigma^2} \right] \mathrm{cos}[\omega_0 (t-t_0)],
\end{equation}
with the half-width-at-half-maximum $\sigma = 5$ fs and the center of the envelope at $t_0 = 18$ fs. The pulse was polarized in the $z-$direction. The maximum intensity is given by $I_\mathrm{max}=c \epsilon_0 E_0^2$, where $c$ is the speed of light, $\epsilon_0$ is the permittivity in vacuum, and $E_0$ is the maximum field strength.

For each frequency and intensity, the  RT-TDDFT-EMD simulations were initiated from the optimized geometry of the H$_2$ molecule adsorbed on the nanoshell, while the initial atomic velocities correspond to an initial temperature of 300~K.
All the atoms were allowed to move freely without any geometry constraint (i.e., no frozen atoms) during the dynamics. The unit cell size in the EMD simulations was changed to $20\times20\times30$ \AA$^3$ to allow for possible desorption of the molecule.

In the calculations with the ghost-augmented basis set (Section~\ref{sec:correct}), the same DZVP basis set and GTH pseudopotentials were used for Ag and Ag ghost atoms. We tested the ghost-augmented basis set on a smaller system, an H$_2$ molecule on a silver chain Ag$_6$. We obtained the same number of emitted electrons from the wire and from H$_2$ for one and two layers of ghost atoms around the wire. Placing the layer of ghost atoms at different distances from the system (from 3 to 10~{\AA}) did not affect the results in any significant way as long as there was an overlap between the basis-set functions of the system and Ag$_\mathrm{g}$. Diffuse basis-set functions aug-cc-Q were also tested and led to the same results, again, if the basis-set functions overlapped.

Data analysis and visualization were performed using Bader analysis,~\cite{Tang2009} NumPy,~\cite{Harris2020} Matplotlib,~\cite{Hunter2007} VESTA,~\cite{Momma2011} and Gnuplot.~\cite{Gnuplot_5.4}

\section{Results and Discussion}
\label{sec:Results}

\subsection{Field intensity and frequency dependence of H$_2$ dissociation on Ag$_{55}^{L1}$}\label{sec:wrong}

In this section, we show the results obtained by employing the RT-TDDFT-EMD methodology following the common practice of using strong external field conditions to speed up the simulations. After analyzing these results, it will become evident that the use of such strong fields can be masking any possible plasmon-induced effect, casting hence doubts on the adequacy of such simulation conditions to interpret experiments on plasmon-enhanced catalysis.

\begin{figure*}[h!]
		\centering
		\includegraphics[width=0.97\linewidth]{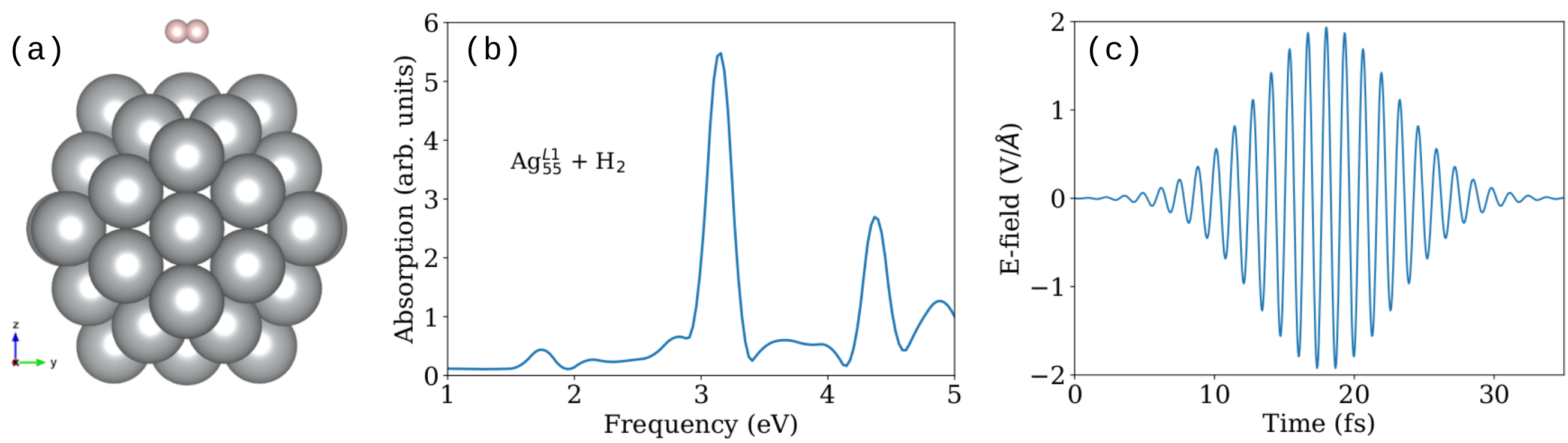}
		\caption{(a) Atomic structure of the relaxed Ag$_{55}^{L1}$ nanoshell with H$_2$ (interatomic distance of 0.75~{\AA}) at a distance of 3~{\AA} from the nanoshell facet. (b) Absorption spectrum of Ag$_{55}^{L1}+$H$_2$. (c) Time-dependent field strength of the external field pulse with a Gaussian envelope ($\sigma = 5$ fs, $t_0 = 18$ fs) with $\hbar\omega_0 = \hbar\omega_\textrm{p} =3.15$ eV. The maximum field strength $E_0 =1.94$ V/{\AA} (0.038 a.u.) corresponds to the maximum intensity $I_\mathrm{max} = 1\times 10^{14}$ W/cm$^2$.}
		\label{fgr:str_abs_spec}
	\end{figure*}

The structure and the RT-TDDFT absorption spectrum of the Ag$_{55}^{L1}+$H$_2$ system are shown in Figure \ref{fgr:str_abs_spec} (a) and (b), respectively. The spectrum exhibits two absorption peaks due to the hybridization of plasmon modes on the inner and outer surfaces of the nanoshell, leading to bonding and anti-bonding resonances. \cite{Prodan2004} The external $z$-polarized field is modeled as a Gaussian pulse centered at $t_0=18$~fs and with half-width-at-half-maximum $\sigma=5$~fs (Figure \ref{fgr:str_abs_spec}(c)). Three field frequencies $\omega_0$ that include the resonant plasmon frequency corresponding to $\hbar\omega_0 =\hbar \omega_\textrm{p}=3.15$~eV and two frequencies in the minima of the absorption spectrum plotted in Figure~\ref{fgr:str_abs_spec}~(b), $\hbar\omega_0 =2$~eV and $\hbar\omega_0 =4.1$~eV, are selected to study the $\omega_0$-dependence of H$_2$ dissociation on Ag$_{55}^{L1}$ and, more specifically, the plasmon role in activating that process. 
The dependence on the field intensity $I$ is also analyzed by considering the two following values for the maximum intensity of the Gaussian pulse, $I_\mathrm{max} = 2\times 10^{13}$ and $1\times 10^{14}$ W/cm$^2$, that agree well with the usual intensities employed in previous studies of this kind.~\cite{Yan2015,YuZhang2018,Hull2020,Hull2023,YiminZhang2021,zhang22,Herring2023}
Note that such high intensities are usually justified by the high computational cost of RT-TDDFT-EMD simulations, which only permit the calculation of the system dynamics for a few tens of fs. 
In contrast, in experiments, the employed field intensities are usually much lower (below 1 MW/cm$^2$) and it can take seconds to observe any meaningful change in the catalytic reaction rate. \cite{Mukherjee2012} Furthermore, it is worthy to remark that the usual physical quantities measured in experiments are the reaction rates and reaction probabilities. These are also the values calculated when using other  computational methods, such as DFT molecular dynamics simulations, by means of a statistically meaningful sampling of the system's initial conditions. However, the purpose of this work is not to calculate the dissociation probability but to clarify the precise dissociation mechanisms and their dependence on the external field conditions, for which it is enough to focus on single dynamical events. To this aim, we rely on the advanced RT-TDDFT-EMD methodology, as it has also been done in similar studies.

\begin{figure*}[h]
 \includegraphics[width=0.99\linewidth]{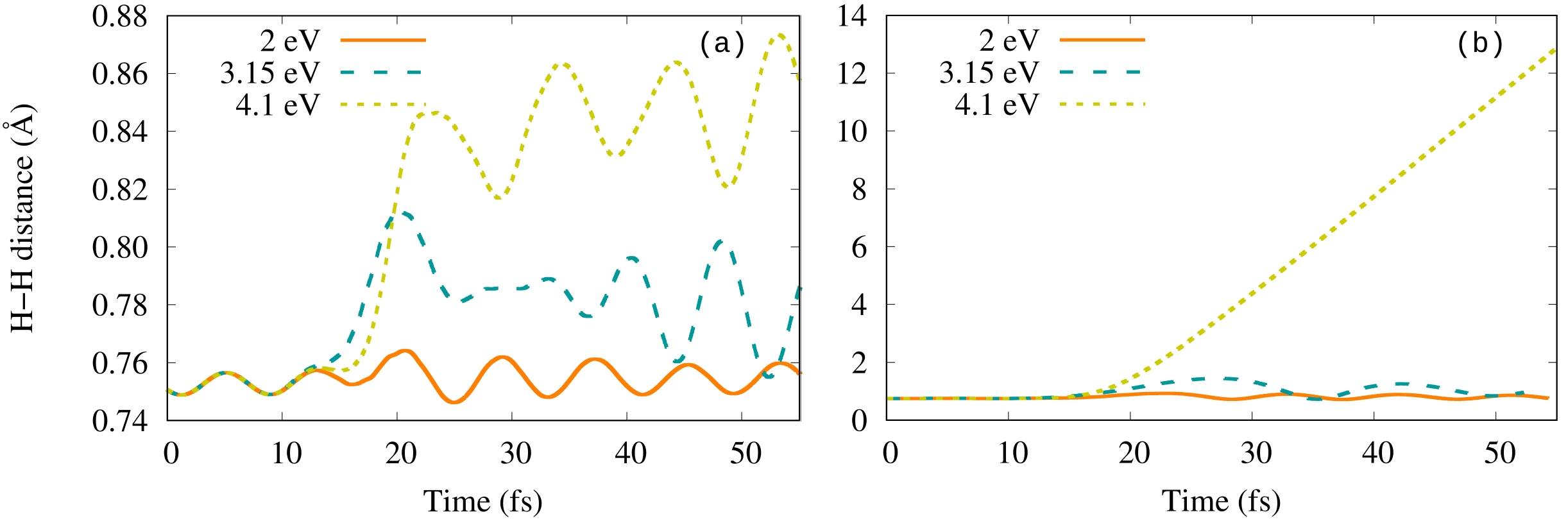}
		\caption{H-H bond length as a function of time for the three chosen field frequencies. Field intensity is (a) $I_\mathrm{max} = 2\times 10^{13}$ W/cm$^2$ and (b) $I_\mathrm{max} = 1\times 10^{14}$ W/cm$^2$. The maximum of the external field arrives at 18~fs.}
		\label{fgr:HH_noghost}
	\end{figure*}

Figure~\ref{fgr:HH_noghost} shows the time evolution of the H-H bond length for the different field conditions under
consideration. At low $I_\textrm{max}$ (panel (a)), the H-H distance oscillates and even deviates gradually from its equilibrium bond length as $\omega_0$ increases, but without dissociating. At high $I_\textrm{max}$ (panel (b)), the bond-length oscillations are more pronounced and at $\hbar\omega_0=4.1$~eV, the molecule dissociates and even desorbs, as shown in Figure~S1~\dag. The $I$-dependence observed here is consistent with the threshold intensity for dissociation observed previously in similar studies. \cite{Yan2015,YuZhang2018,Hull2020,Hull2023,YiminZhang2021,zhang22} The behavior at different frequencies, however, is rather puzzling. The fact that H$_2$ does not dissociate at $\omega_\textrm{p}$ but at a higher frequency regardless of lying in a minimum of the absorption spectra casts doubts on the influence of the plasmon excitation in activating the dissociation in this system at the considered field intensities. Interestingly, the obtained $\omega_0$-dependence contrasts with the results for H$_2$ on Ag$_6$, showing that dissociation occurs at $\omega_0=\omega_\mathrm{p}$ but not at other frequencies.~\cite{Yan2015} Unfortunately, there is no more information on other systems that could 
clarify the actual role of the plasmon excitation in activating reactions under strong-field conditions, since most of the existing studies only explore resonant field frequencies (plasmon resonance and other maxima in the absorption spectrum related to interband transitions, for instance).

\begin{figure*}[ht!]
		\includegraphics[width=0.99\linewidth]{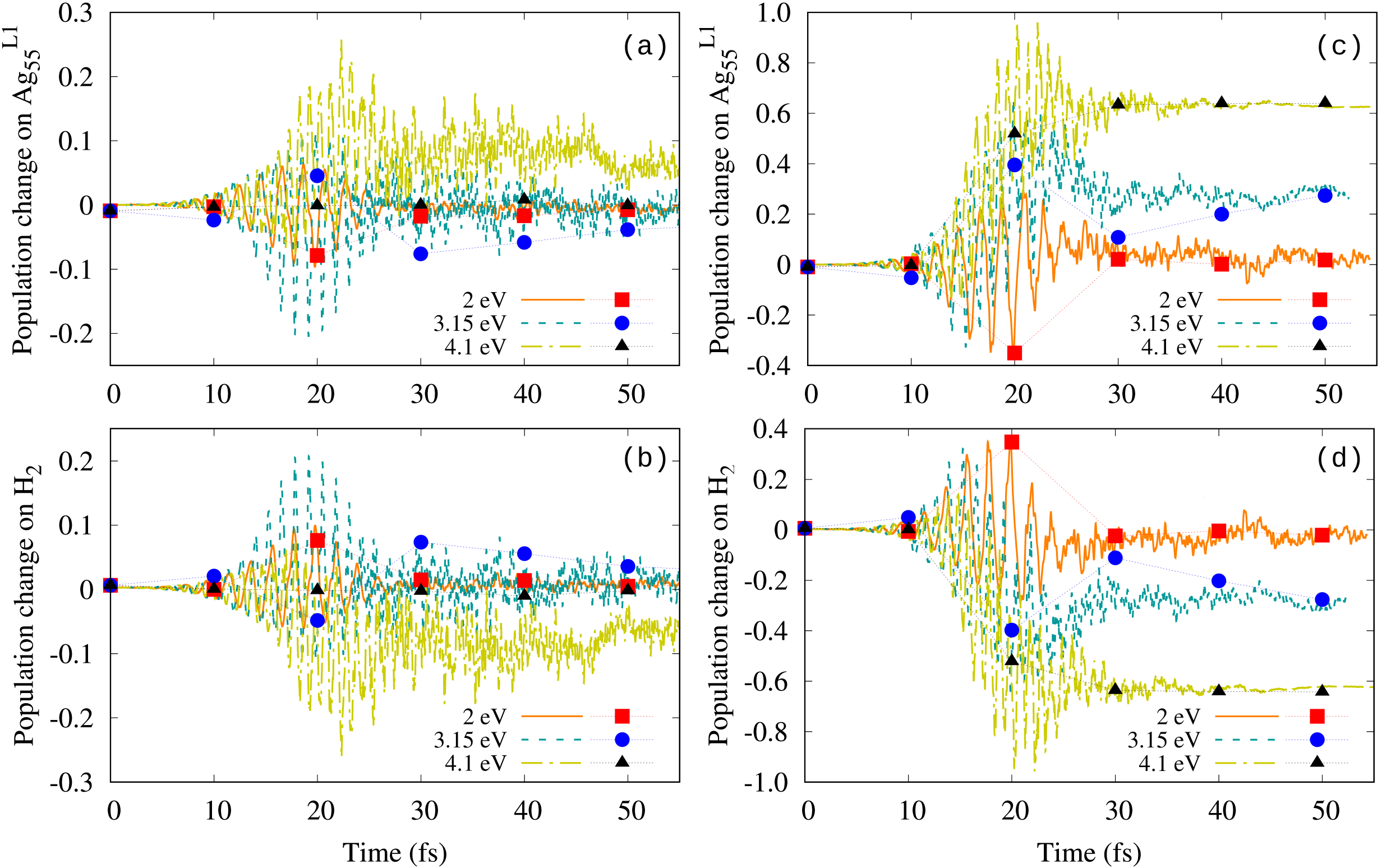}
		\caption{Time evolution of the Mulliken (lines) and Bader (symbols) population change [$\Delta N\mathrm{e} = N_\mathrm{e}(t) - N_\mathrm{e}(t=0)$] on (a,c) Ag$_{55}^{L1}$ and (b,d) H$_2$ for the three studied field frequencies. Field intensity is (a,b) $I_\mathrm{max} = 2\times 10^{13}$ W/cm$^2$ and (c,d) $I_\mathrm{max} = 1\times 10^{14}$ W/cm$^2$.}
		\label{fgr:charge_noghost}
	\end{figure*}

The analysis of both the Mulliken population and the Bader distribution provides information on the transient electron transfer between the molecule and the nanoparticle that is caused by the external field. 
Figure \ref{fgr:charge_noghost} shows the transient change in the Mulliken population 
on both the nanoshell and the molecule
for $I_\mathrm{max} = 2\times 10^{13}$ W/cm$^2$ and $I_\mathrm{max} = 1\times 10^{14}$ W/cm$^2$, respectively. The change is calculated in each case as the difference with respect to the value at $t=0$, i.e., $\Delta N_\mathrm{e}(t)= N_\mathrm{e}(t)-N_\mathrm{e}(t=0)$.
Hence, negative values of $\Delta N_\mathrm{e}$ mean a reduction of the number of electrons. 
Figure~\ref{fgr:charge_noghost}(a,b) shows that at low $I_\textrm{max}$ and field frequencies $\hbar\omega_0=2.0$ and 3.15~eV, the electron distribution oscillates between the nanoshell and the molecule following the external field and it progressively recovers its initial value once the external field has been switched off.  
The larger amplitude of the oscillations at the plasmon frequency $\hbar\omega_0=3.15$~eV compared to 2~eV seems to be consistent with the plasmonic resonance influence. Similar back and forth charge oscillations between the metal nanoparticle and the molecule for the duration of the pulse were obtained in other systems, such as O$_2$ and N$_2$ on Au nanocube
\cite{Herring2023} and H$_2$ on AuNP. \cite{YuZhang2018} 
However, the behavior at 4.1~eV is rather odd. First, the maximum of the oscillation amplitude is shifted toward later time comparing to the maximum of the external field (18 fs). Furthermore, after the field has been switched off, the charge on the molecule is positive (the number of electrons in the molecule is reduced), which actually indicates the transfer of electrons from the molecule to the nanoshell. 
 
Although not much discussed, a similar reduction of the number of electrons on the molecule upon switching off the external field has also been observed in various systems. Yan et al. \cite{Yan2015} showed that the number of electrons on H$_2$ adsorbed on Ag$_6$ starts to decrease after passing the maximum of the external pulse and the reduction amounts to 1 electron once the pulse is off at field strength 2.5 V/{\AA} ($I_\mathrm{max} = 1.656 \times 10^{14}$ W/cm$^2$). 
Similar observations were made in refs \citenum{Herring2023,YuZhang2018}, where the charge change on the molecule after the pulse was switched off was different from zero and either positive or negative depending on the molecule (reduction of the number of electrons on H$_2$, \cite{YuZhang2018} slight reduction of the number of electrons on O$_2$, and a slight increase on N$_2$. \cite{Herring2023})
Kuda-Singappulige et al. \cite{KudaSingappulige2023} analyzed the Mulliken population on O$_2$ activated on Ag$_8$, which revealed the transfer of electrons from the oxygen molecule to the silver nanoparticle in all the dissociative cases.
From the examples listed above, it is evident that the dissociation mechanism is system-specific. Not only electron transfer from the NP to the molecule can lead to dissociation, as is commonly accepted, but also the transfer of electrons from the molecule to the NP.
Moreover, Herring et al. suggest that the charge transfer is neither necessary nor sufficient for the dissociation to occur. \cite{Herring2023}

At the higher field intensity $1\times 10^{14}$ W/cm$^2$, the largest charge fluctuations also occur at the largest field frequency of 4.1 eV (Figure \ref{fgr:charge_noghost}(c,d)). At $t\simeq20$ fs, when the molecule starts to dissociate (see Figure \ref{fgr:HH_noghost}~(b)), the transient positive charge on H$_2$ at this frequency corresponds to losing about one electron.
A transient loss of about 0.6 electrons on H$_2$ is also observed at the plasmon frequency, however it seems to be insufficient to cause dissociation. For completeness, we also calculated the Bader distribution every 10 fs along the simulation. Similar population changes were obtained by both Mulliken and Bader analysis.

\begin{figure*}[h!]
 \centering
		\includegraphics[width=0.98\linewidth]{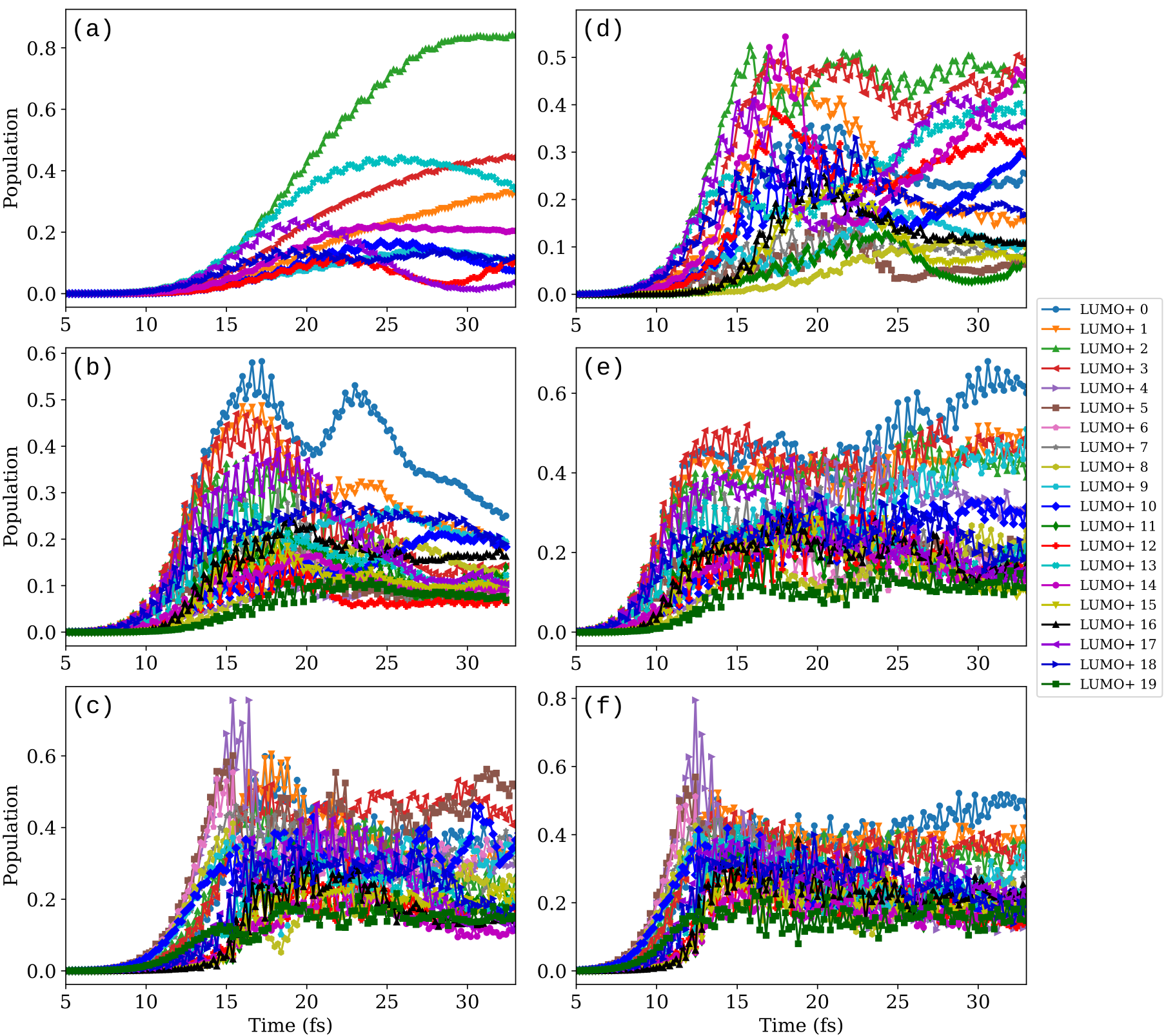}
		\caption{ Time evolution of the Ag$_{55}^{L1}$+H$_2$ orbital populations induced by an external field with intensity (left panels, (a,b,c)) $I_\mathrm{max} = 2\times 10^{13}$ W/cm$^2$ and  (right panels, (d,e,f)) $I_\mathrm{max} = 1\times 10^{14}$ W/cm$^2$. For each $I_\mathrm{max}$, the field frequency is: (a,d) $\hbar \omega_0 =$ 2 eV, (b,e) $\hbar\omega_0 =$ 3.15 eV, and (c,f) $\hbar\omega_0 =$ 4.1 eV. Orbital populations are calculated every 0.2 fs as sums of the squares of the projections of the time-dependent occupied MOs on the initially unoccupied orbitals. Only populations with maximum values $> 0.1$ are plotted.}
		\label{fgr:projections}
	\end{figure*}

The proposed mechanism of bond stretching (or dissociation) 
by electron transfer from the metallic nanoparticle to the molecule is often supported by analyzing the time-dependent occupation of the initially unoccupied molecular orbitals (MOs). \cite{YiminZhang2021,zhang22,Hull2020} Figure \ref{fgr:projections} shows the orbital populations calculated from the
projections of the time-dependent occupied MOs of Ag$_{55}^{L1}$+H$_2$ on the initially unoccupied MOs. For each initially unoccupied orbital $\psi_{\mathrm{LUMO+}n}$, where $n$ runs from 0 to 50, the population $P_{\mathrm{LUMO+}n}$ at time $t$ is calculated as $P_{\mathrm{LUMO+}n}(t)=\sum_{i=1}^{N_{\mathrm{occ}}} |\langle\psi_{\mathrm{LUMO+}n}|\psi_i(t)\rangle|^2$, where $\psi_i(t)$ are the time-dependent occupied orbitals. Note that the projections are calculated from the propagation of the electronic states only (i.e., fixing the nuclei at their equilibrium positions). Thus, the information they provide will be meaningful as long as the system geometry is not strongly perturbed.

Finite populations of initially unoccupied MOs of very high energies
are observed at all frequencies and both intensities.  The number of high-energy MOs with a sizable population increases with both the field intensity and frequency, making it more and more difficult to distinguish among the different projection curves. 
Analysis of the spatial distribution of each unoccupied MO shows that out of 19 excited states, only 8 have features on the H$_2$ molecule (see Figure~S2~\dag). Orbitals LUMO+3, +4, +10, +13, and +18 have a bonding, while LUMO+6, +16, and +19 have an antibonding character on H$_2$. Notably, the populations on LUMO+6, +16, and +19 are relatively high at the frequency 4.1 eV at both intensities, 
which can explain the large internuclear oscillations and dissociation of H$_2$ observed at this frequency in Figure~\ref{fgr:HH_noghost}(a,b). None of the three antibonding orbitals are populated at 2 eV and $I_\mathrm{max} = 2\times 10^{13}$ W/cm$^2$, which is reflected in the H-H bond evolution in this case showing no activation of the molecular bond (Figure~\ref{fgr:HH_noghost}(a)). At 2 eV and $I_\mathrm{max} = 1\times 10^{14}$ W/cm$^2$, orbital LUMO+16 is populated, giving rise to a slight bond activation (reaching 0.93 \AA ~at 23 fs). At the LSPR frequency 3.15 eV, all three MOs with the antibonding features on H$_2$ have finite populations, which are however not enough to dissociate the molecule. 
Sizable populations 
on high-energy unoccupied MOs were also observed by other authors, for instance, for CO$_2$ on Ag$_6$~\cite{YiminZhang2021} (LUMO$+10$) and on Ag$_{20}$~\cite{Zhang2021b} (up to LUMO$+12$), for NH$_3$ on Ag$_6$~\cite{zhang22} (up to LUMO$+13$)
and for N$_2$ on Ag$_8$~\cite{Hull2020} (up to LUMO$+9$).
Overall, the finite populations of MOs up to LUMO+\textcolor{red}{19} 
indicate that electrons are in a highly excited state and that there is no apparent feature that would distinguish the resonant frequency of 3.15~eV at the strong fields considered.

\begin{figure*}[ht]
  \centering
		\includegraphics[width=0.99\linewidth]{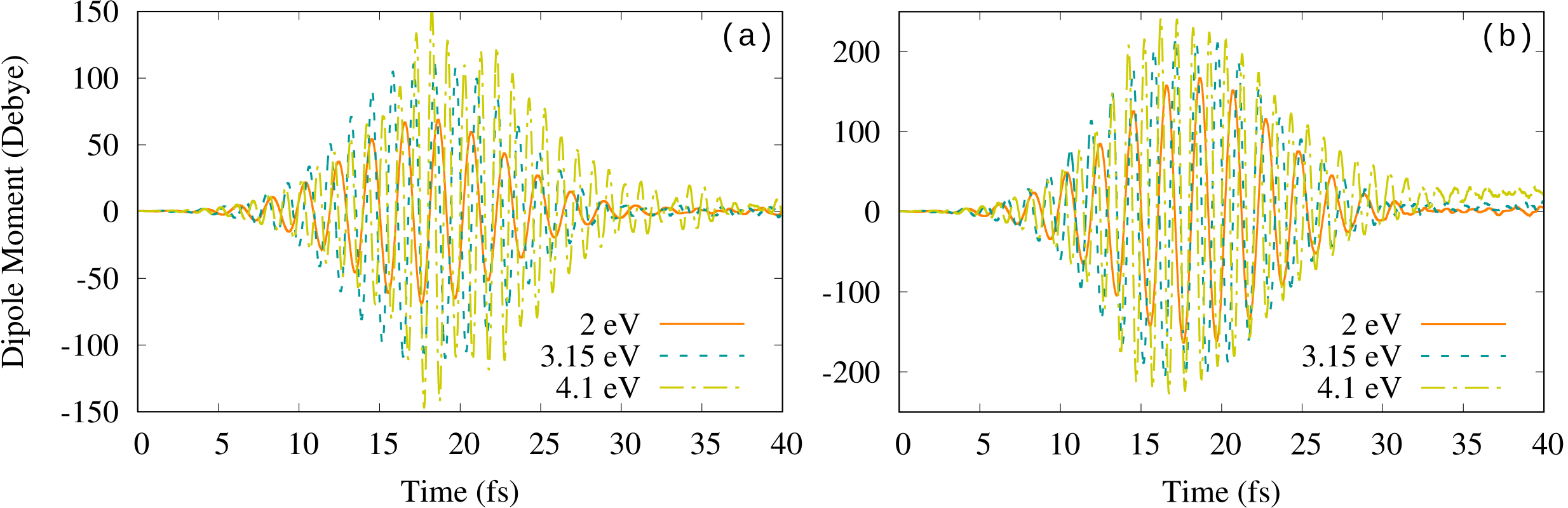}
		\caption{Time-dependent electric dipole moment for (a) $I_\mathrm{max} = 2\times 10^{13}$ W/cm$^2$ and (b) $I_\mathrm{max} = 1\times 10^{14}$ W/cm$^2$.}
		\label{fig:dipole}
	\end{figure*}

The occupation of high-energy MOs and the fact that $\hbar \omega_0 = 4.1$ eV (minimum in the absorption spectrum) gives us a larger bond separation than the plasmon frequency 3.15 eV suggest 
the nonlinearity of the observed processes. The comparative analysis of the field-induced dipole moment and its dependence on the external field properties ($I_\textrm{max}$ and $\omega_0$) allows us to further confirm the existence of nonlinear effects at these strong fields. Indeed, as shown in Figure \ref{fig:dipole} for both intensities, the dipole moment oscillations are larger at 4.1 eV than at the resonant plasmon frequency of 3.15 eV at which the induced dipole is expected to be the largest. 
Fourier transform of the induced dipole shows that the amplitude is larger at 4.1 eV than at 3.15 eV (see Figure~S4~\dag). It also shows that high harmonics are excited at the three frequencies considered in this work (see Figure~S5~\dag). A strong external field may lead to electron emission and ionization, which, however, are not discussed in the RT-TDDFT-EMD studies cited in this section. In what follows, we show that such nonlinear effects, induced by the strong external field, have to be taken into account for a correct interpretation of the RT-TDDFT-EMD results. In particular, it is important to realize that processes such as ionization are not correctly described in RT-TDDFT-EMD simulations that, as done in this section and by other authors,~\cite{Hull2020,Hull2023,KudaSingappulige2023} use atom-centered basis sets, which can not describe the continuum. Thus, in order to incorporate these missing excitations in our simulations, we have repeated all the calculations adding the so-called floating centers (or ``ghost" atoms) \cite{Soriano2008,White2015} around our system. The new results and the consequences of such an improvement are discussed in the next section.

\subsection{Strong-field effects with augmented basis set}\label{sec:correct}

To account for possible electron emission processes, we improve the basis set by adding an additional layer of 92 silver ghost atoms (Ag$_\mathrm{g}$) around the nanoshell (Figure~S6~\dag). The ghost layer corresponds to the shell number 4 of the icosahedral cluster. The ghost atoms have no physical characteristics (no nuclear charge and no electrons) and only serve for placing basis functions in the empty space outside the cluster to model electronic unbound states. The convergence of the results with the number of ghost atoms and their distance to the system was tested on a smaller system (see details in the methodology section ``Ehrenfest molecular dynamics simulations").

\begin{figure*}[ht]
 \includegraphics[width=0.99\linewidth]{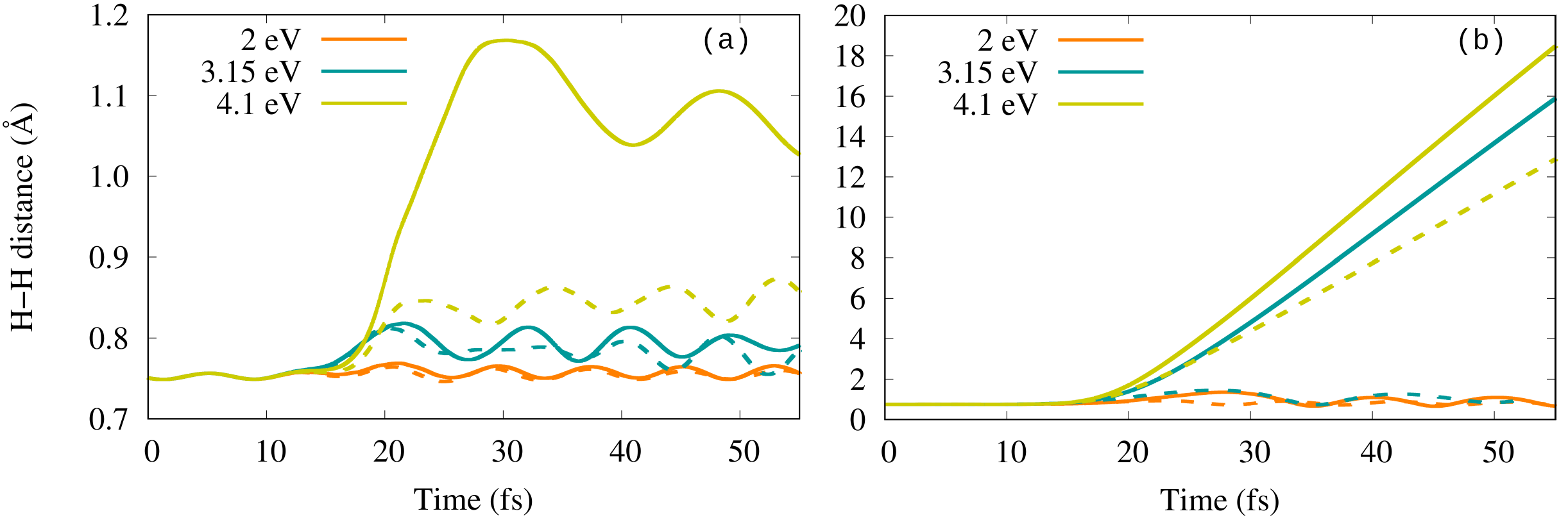}
		\caption{Comparison of the H-H bond length with (solid lines) and without (dashed lines) augmented basis. Field intensity is (a) $I_\mathrm{max} = 2\times 10^{13}$ W/cm$^2$ and (b) $I_\mathrm{max} = 1\times 10^{14}$ W/cm$^2$.}
		\label{fig:HH-ghost-vs-noghost}
	\end{figure*}

The time evolution of the H-H bond length obtained in the calculations with and without the additional basis-set 
functions are compared in Figure \ref{fig:HH-ghost-vs-noghost} for each field frequency and intensity. At low $I_\textrm{max}$, Figure \ref{fig:HH-ghost-vs-noghost}(a) shows that H$_2$ dissociation is not observed with the ghost-augmented basis set either. For the lower frequencies (2.0 and 3.15~eV), the results are rather independent of the basis set.  
However, at the highest frequency considered ($\hbar\omega_0=4.1$~eV), the difference is substantial. With the ghost-augmented basis set, the H-H bond stretches to approximately 1.1~{\AA}, as compared to 0.85~{\AA} without it. At high $I_\textrm{max}$ (Figure \ref{fig:HH-ghost-vs-noghost}(b)), the molecule dissociates at 3.15 eV and 4.1 eV when using the ghost-augmented basis set, while there was no dissociation at 3.15~eV when no ghost atoms were included.
    \begin{figure*}[h!]
    \includegraphics[width=0.98\linewidth]{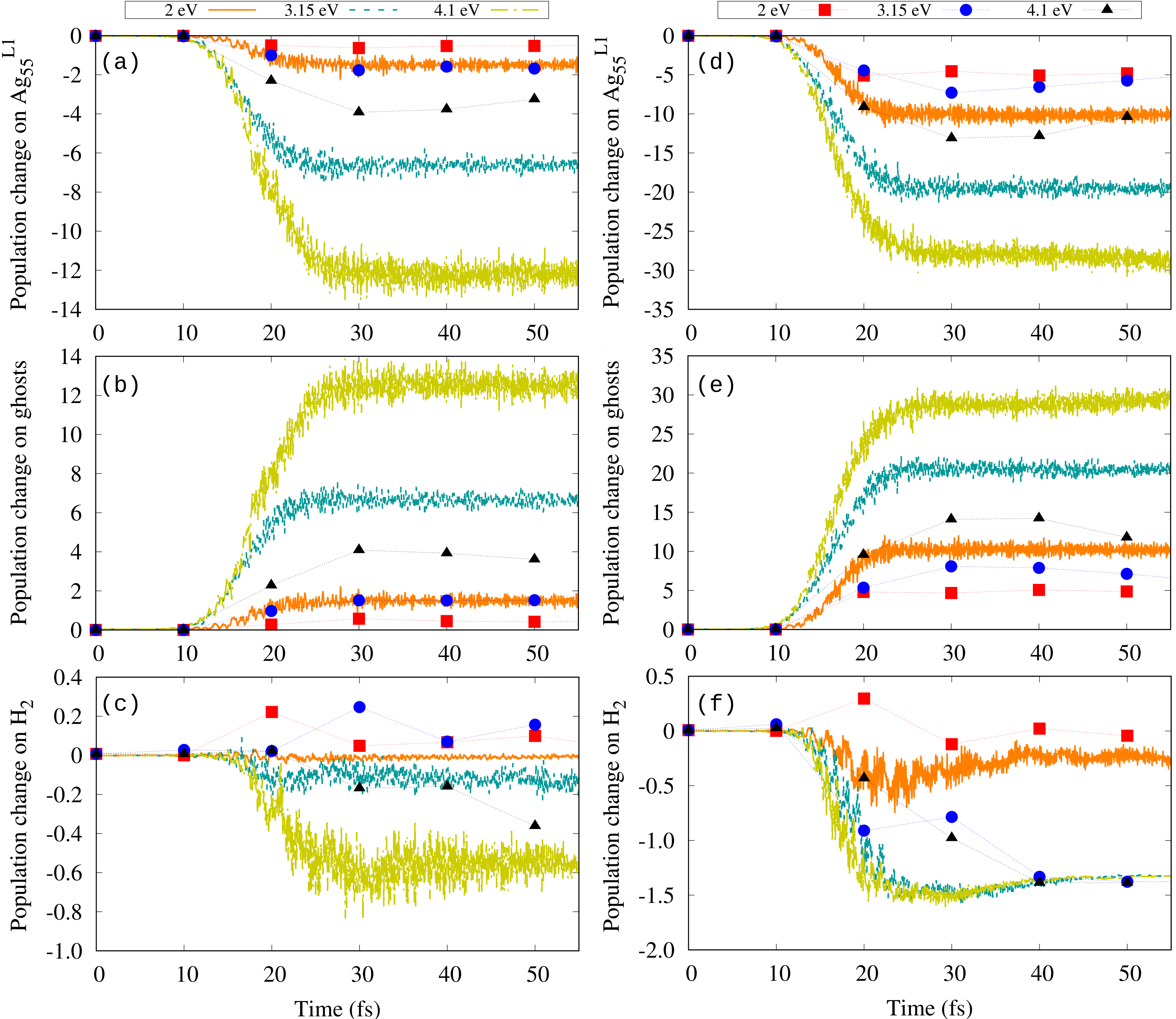}
		\caption{Time evolution of the Mulliken (lines) and Bader (symbols) population change [$\Delta N_\textrm{e} = N_\textrm{e}(t) - N_\textrm{e}(t=0)$]
        on (a) Ag$_{55}^{L1}$, (b) ghost atoms, and (c) H$_2$ for $I_\mathrm{max} = 2\times 10^{13}$ W/cm$^2$, and on (d) Ag$_{55}^{L1}$, (e) ghost atoms, and (f) H$_2$ for $I_\mathrm{max} = 1\times 10^{14}$ W/cm$^2$.}
		\label{fig:Mulliken-ghost-1e13}
	\end{figure*}
 
To understand the difference in dissociation observed in Figure~\ref{fig:HH-ghost-vs-noghost}, we analyzed the induced charge on the Ag nanoshell, the H$_2$ molecule, and the ghost shell, separately. Figure \ref{fig:Mulliken-ghost-1e13} (a)-(c) shows the Mulliken population change over time for $I_\mathrm{max} = 2\times 10^{13}$~W/cm$^2$.  
In contrast to the results without additional basis, both the nanoshell and the molecule lose electrons, which is manifested as the transfer of electrons to the ghost atoms. This is also evident from the populations of the initially unoccupied orbitals as shown in Figure~S3{~\dag} for 3.15 eV and both intensities for the cases without and with the ghost atoms. In the case with the ghost atoms, additional orbitals up to LUMO+48 have populations higher than 0.1. 
The transfer of electrons to the ghost atoms increases with increasing frequency of the external field (Figures~\ref{fig:Mulliken-ghost-1e13}(b,e)). The obtained $\omega_0$-dependence of $\Delta N_\textrm{e}(t)$ 
suggests that the electron loss (and subsequently the H-H bond length) is not much related to any plasmon effect.
Bader analysis overall shows smaller ionization, but nevertheless leads to the same conclusion, i.e., that the molecular bond is activated more at 4.1 eV because H$_2$ becomes positively charged.

At high $I_\textrm{max}$ (Figure \ref{fig:Mulliken-ghost-1e13}(d)-(f)), both the Mulliken and Bader population change show that at $\hbar\omega_0=3.15$ and 4.1~eV, the H$_2$ molecule loses approximately 1-1.5 electrons in the time interval at which dissociation takes place. Thus, it is the ionization of the molecule that promotes the bond weakening and its subsequent dissociation for both plasmonic and out-of-resonance frequencies. This observation has important implications for the established way of modeling plasmonic catalysis from first principles using strong external fields.

A single-photon absorption may not be responsible for the ionization of our system. 
The ionization potential of the Ag$_{55}^{L1}+$H$_2$ system obtained from the difference of the DFT total energies $\mathcal{E}$ for the charged and neutral system, $I_p = \mathcal{E}$ (Ag$_{55}^{L1}+$H$_2)^+ - \mathcal{E} ($Ag$_{55}^{L1}+$H$_2)$ is 3.64 eV. 
A possible explanation for the electron loss is thus multiphoton absorption and subsequent ionization or electron tunneling due to suppression of the potential barrier by the strong external field (above-threshold ionization). \cite{M_Protopapas_1997} 
To assess which of these processes prevails, we estimate the Keldysh parameter $\gamma$ for our field conditions. \cite{Fennel2010,Li2014} Keldysh parameter is defined as $\gamma = \sqrt{I_p/(2U_p)}$, where $I_p$ is the ionization potential, $U_p = E_0^2/(4 \omega_0^2)$ is the ponderomotive potential, $E_0$ is the field strength, and $\omega_0$ is the field frequency (all expressions are in atomic units, a.u.). Tunneling ionization dominates when $\gamma < 1$, while multiphoton ionization is the dominating mechanism when $\gamma > 1$. 
Using the value $I_\mathrm{p} = 3.64$ eV, at $I_\mathrm{max} = 1\times 10^{14}$ W/cm$^2$, we obtain $\gamma= 1.58$ for $\hbar \omega_0 =$ 3.15 eV and $\gamma=2.07$ for $\hbar \omega_0 =$ 4.1 eV, meaning that the multiphoton ionization dominates. Note also that these estimations are made for the emission of a single electron, whereas, as shown in Figure~\ref{fig:Mulliken-ghost-1e13}, several electrons are emitted. The energy threshold for multiple electron emission is larger than $I_p$, which implies larger values of the corresponding Keldysh parameter, supporting the multiphoton character of the process.

Our results regarding the ionization of H$_2$ and Ag$_{55}^{L1}$ are in line with experimental findings. Dissociative ionization of gas-phase H$_2$ has been observed experimentally at similar external field frequency and intensity. \cite{Lopez2017} 
The ionization of the H$_2$ molecule on the Ag$_{55}^{L1}$ nanoshell surface is facilitated because the ionization potential of this system is much lower than that of the H$_2$ molecule in vacuum due to level hybridization between Ag and H (see projected density of states (PDOS) in Figures~S7 and S8~\dag). The calculated DFT ionization potential of the isolated H$_2$ is 14.1 eV, while it is 3.64 eV for Ag$_{55}^{L1}$+H$_2$.
Experimentally, a dissociative above-threshold double ionization of H$_2$ after absorbing more than 10 photons has been observed at near-infrared pulse intensity of the order of $10^{14}$ W/cm$^2$. \cite{Pan2021} 
Ionization probability for single Ag atoms ($I_p = 7.5$ eV \cite{Radzig_1985}) has been estimated to reach 100\% at intensity $2\times 10^{13}$ W/cm$^2$ (neutral Ag atom irradiated by 800 nm (1.55 eV) 35 fs pulse). \cite{Konda2021} 

The fact that the nanoshell loses up to 10 electrons (according to Bader decomposition) may affect its properties. Indeed, our calculations of the absorption spectrum for a charged system $[$Ag$_{55}^{L1}]^{+10}$ show that the plasmon peak shifts to lower energy (see Figure~S9~\dag). As a result, all three frequencies studied here are non-resonant when the nanoshell is ionized. It is important to emphasize that the nonlinear effects induced by a strong external field (such as generation of higher harmonics and multiphoton processes) dominate over plasmonic effects, which is why using high-intensity field pulses when modeling plasmonic catalysis requires careful consideration.

\section{Conclusions}
\label{sec:Conclusion}
In this work, we apply RT-TDDFT combined with Ehrenfest dynamics to investigate the effects of external field intensity and frequency on the dissociation of H$_2$ on the surface of the Ag$_{55}^{L1}$ nanoshell. First, by resorting to the methodology and external field conditions used in several similar studies, we observe no
molecular dissociation at the lower intensity considered ($I_\mathrm{max} = 2\times 10^{13}$~W/cm$^2$) and only a slight bond stretching at $\omega_0 \geq \omega_\mathrm{p}$. At the higher intensity ($I_\mathrm{max} = 1\times 10^{14}$ W/cm$^2$), the molecule dissociates at $\omega_0 > \omega_\mathrm{p}$. However, no dissociation is observed at the plasmon frequency $\omega_\mathrm{p}$ at either the low or high intensities. Such a behavior, together with the highly excited state of the system evident from the population analysis of initially unoccupied MOs, indicates nonlinearities of the studied processes. Indeed, analyzing the dipole moments at all frequencies of the external field, we observe a clear manifestation of the nonlinear behavior, namely, the absence of dipolar resonance at $\omega_\mathrm{p}$.

Next, by taking the nonlinearity into account (as opposite to the linear regime in which multiphoton processes do not occur) and by improving the basis-set to represent the continuum, we observe that both H$_2$ and the nanoshell lose electrons. As a result, H$_2$ dissociates at the higher intensity and frequencies $\omega_0 \geq \omega_\mathrm{p}$. Assessing the external field conditions, the Keldysh parameter, and the ionization potential of our system, we suggest that the dissociation is caused by the multiphoton absorption and subsequent ionization. No dissociation is observed in non-ionizing cases.

Our study emphasizes that modeling molecular dissociation on plasmonic nanoparticles in a strong external field without taking into account the effects such a field can cause can be misleading. Nonlinear effects induced by a strong external field dominate over plasmonic effects, changing the optical and electronic properties of the system. Thus, it is crucial to consider the implication of applying strong external fields in simulations intended to study plasmonic catalysis. Moreover, the existence of a high-intensity threshold for dissociation in computational studies makes it difficult to extrapolate the results to experiments aimed to investigate plasmon-induced catalysis. Experimental setups typically employ much lower intensities, below 1 MW/cm$^2$. As a result, the strong-field phenomena we observe would not occur in actual experimental conditions.

\section*{Author contributions}
Conceptualization: MA, NEK, JIJ; Methodology: NEK; Calculations: NK, MA;
Data analysis and validation: NEK, MA, JIJ; Visualization: NEK; Writing – original draft preparation: NEK; Writing – review and editing: NEK, MA, JIJ; Funding acquisition: MA, JIJ. All authors have accepted responsibility for the entire content of this manuscript and approved its submission.

\section*{Conflicts of interest}
There are no conflicts to declare.

\section*{Data availability}

 The data generated during the current study are presented in the article and in the ESI~\dag.

\section*{Acknowledgements}

We are grateful to Prof. Andrei G. Borisov from Université Paris-Sud for helpful discussions. Financial support was provided by the
Spanish MCIN$/$AEI$/$10.13039$/$501100011033$/$, FEDER Una manera de hacer Europa (Grant No.~PID2022-140163NB-I00), Gobierno Vasco-UPV/EHU (Project No.~IT1569-22), and the Basque Government Education Departments’ IKUR program, also co-funded by the European NextGenerationEU action through the Spanish Plan de Recuperación, Transformación y Resiliencia (PRTR). We also acknowledge the HPC resources provided by the Donostia International Physics Center (DIPC) Supercomputing Center.




\bibliography{plasmon-induced-hot-carriers} 
\end{document}


\title{Supplementary Information:\\
\bf Strong-field effects in the photo-induced dissociation of the hydrogen molecule on a silver nanoshell}
\author{Natalia E. Koval,$^{\ast}$\textit{$^{a}$} J. I\~{n}aki Juaristi,\textit{$^{bac}$} and Maite Alducin\textit{$^{ac}$}}
\date{} 
\maketitle

$^*$ natalia.koval@ehu.eus\\
\textit{$^{a}$~Centro de F\'{\i}sica de Materiales CFM/MPC (CSIC-UPV/EHU), Paseo Manuel de Lardizabal 5, 20018 Donostia-San Sebasti\'an, Spain; E-mail: natalia.koval@ehu.eus}\\
\textit{$^{b}$~Departamento de Pol\'{i}meros y Materiales Avanzados: F\'{i}sica, Qu\'{i}mica y Tecnolog\'{i}a, Facultad de Qu\'{i}mica (UPV/EHU), Apartado 1072, 20080 Donostia-San Sebasti\'{a}n, Spain }\\
\textit{$^{c}$~Donostia International Physics Center (DIPC),
Paseo Manuel de Lardizabal 4, 20018 Donostia-San Sebasti\'an, Spain }

\vspace{3cm}

\section{Additional figures}

\begin{figure}[h!]
 \centering
		\includegraphics[width=0.99\linewidth]{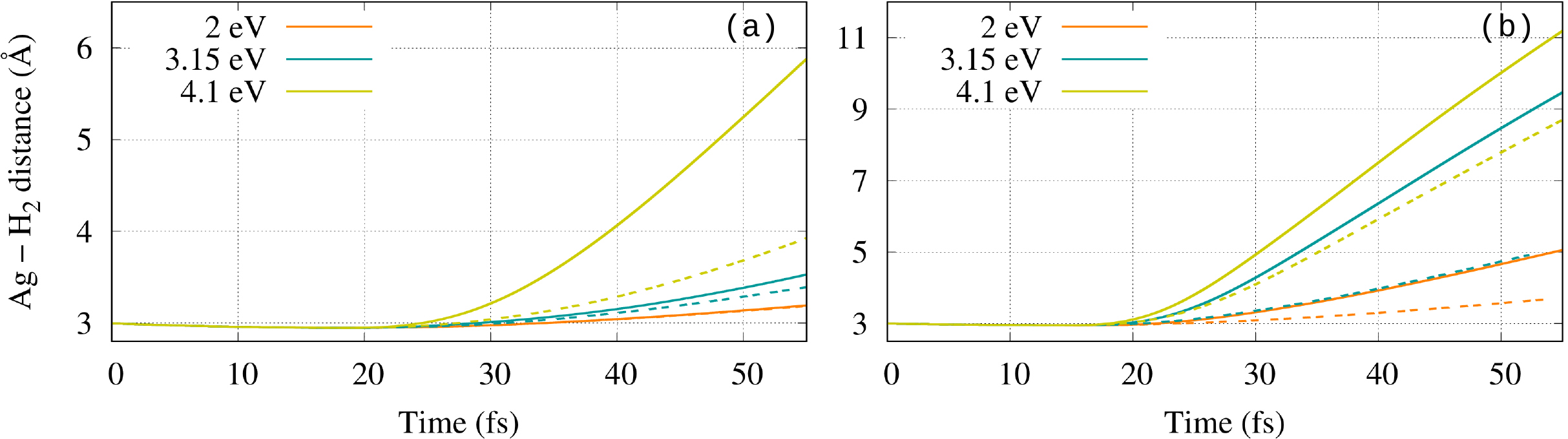}
		\caption{Distance from the top Ag atom to the center of the H$_2$ molecule as a function of time: (a) $I_\mathrm{max} = 2 \times 10^{13}$ W/cm$^2$, (b) $I_\mathrm{max} = 1 \times 10^{14}$ W/cm$^2$. Solid (dashed) lines correspond to the calculations with (without) ghost atoms.}
		\label{fig:dens}
	\end{figure}

  	\begin{figure}
  \centering
    \includegraphics[width=0.99\linewidth]{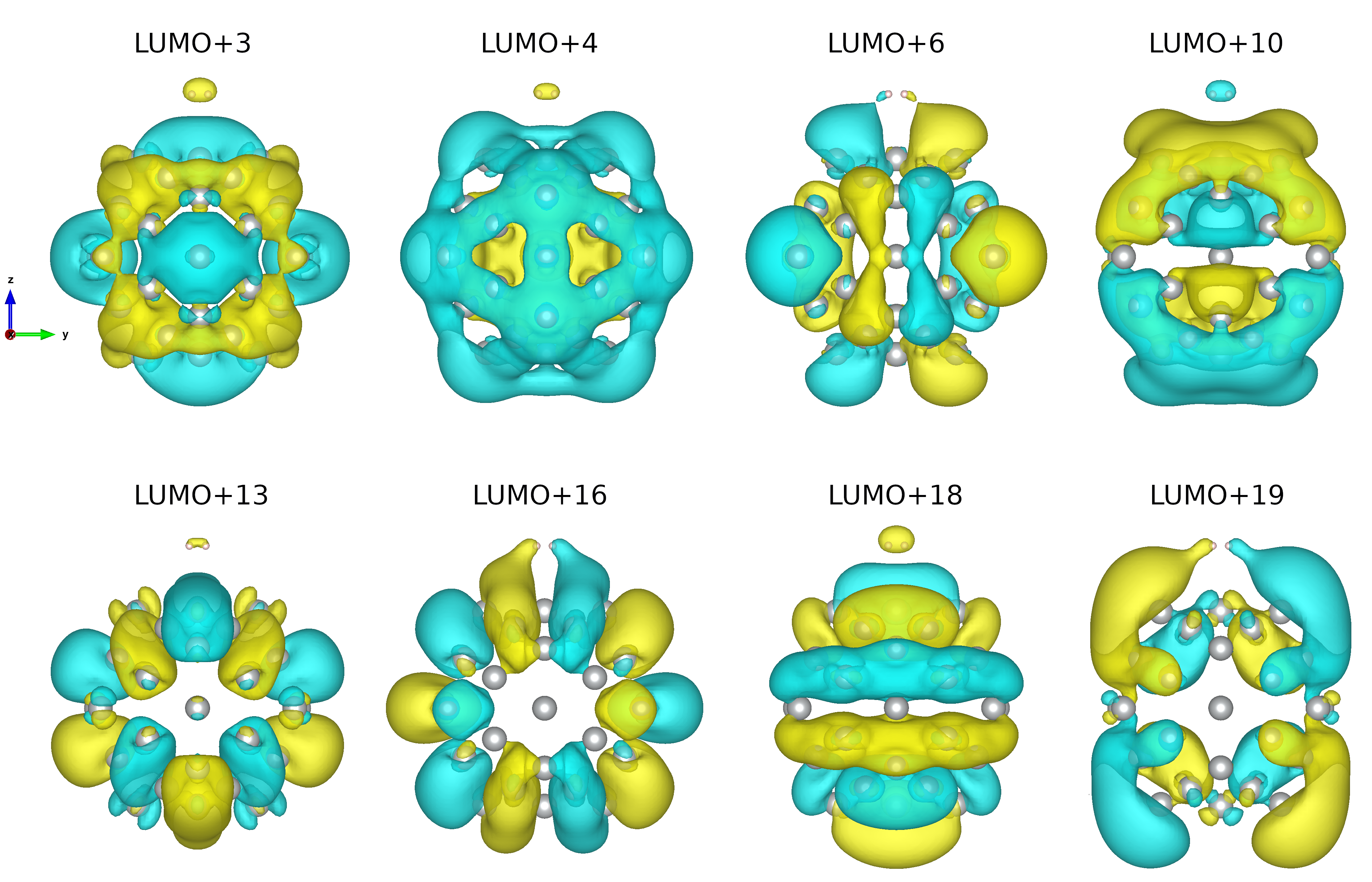}
		\caption{Ground-state unoccupied molecular orbitals of Ag$_{55}^{L1}$+H$_2$. Orbitals LUMO+3, +4, +10, +13, and +18 have a bonding, while LUMO+6, +16, and +19 have an antibonding character on H$_2$.}
		\label{fig:lumo10}
	\end{figure}

  	\begin{figure}
  \centering
    \includegraphics[width=0.99\linewidth]{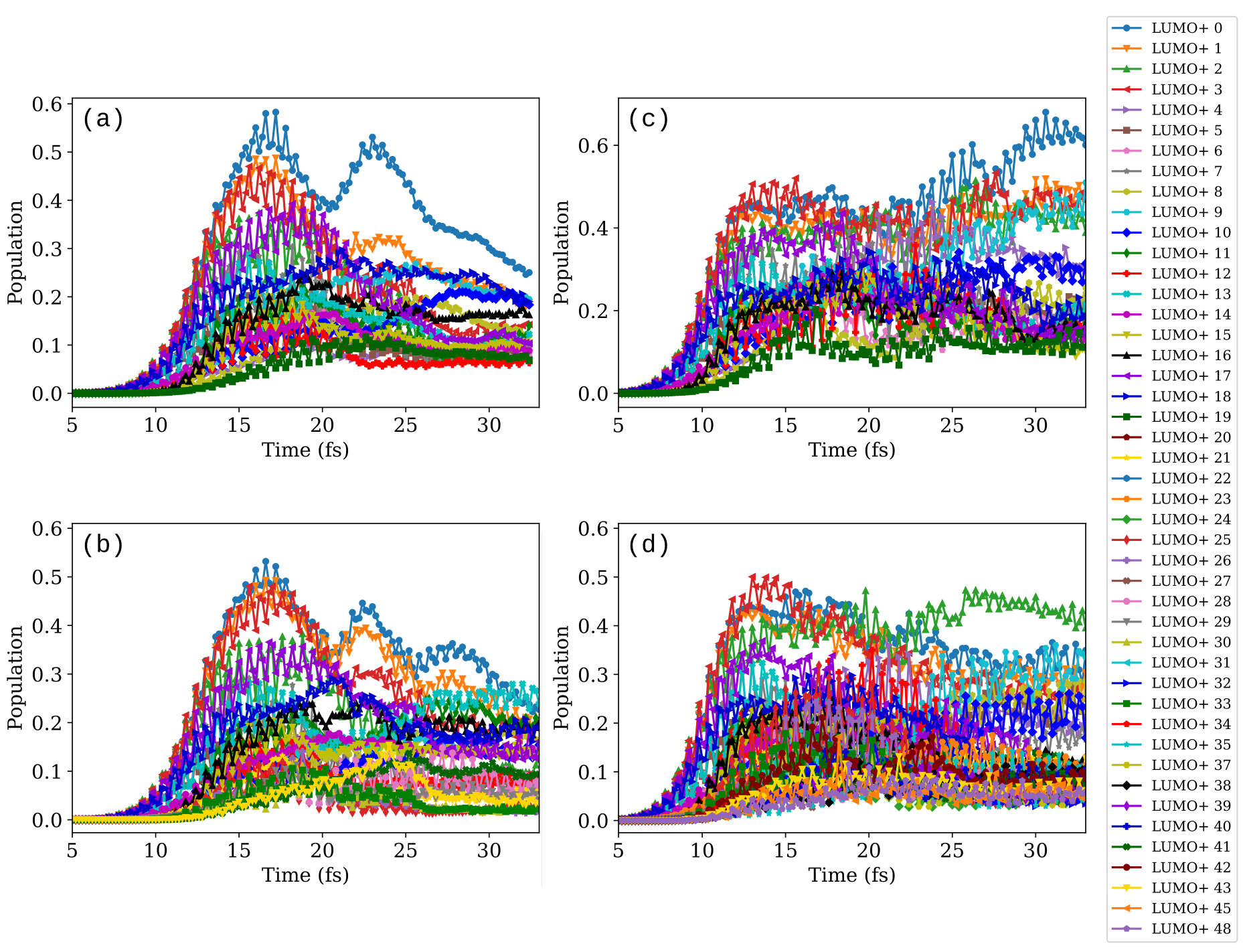}
		\caption{Time evolution of the Ag$_{55}^{L1}$+H$_2$ orbital populations induced by an external field with intensity (left panels, (a,b)) $I_\mathrm{max} = 2\times 10^{13}$ W/cm$^2$ and  (right panels, (c,d)) $I_\mathrm{max} = 1\times 10^{14}$ W/cm$^2$. (a) and (c) correspond to the case without the ghost atoms, while (b) and (d) - with the ghost atoms. The field frequency is $\hbar\omega_0 =$ 3.15 eV. Orbital populations are calculated every 0.2 fs as sums of the squares of the projections of the time-dependent occupied MOs on the initially unoccupied orbitals. Only populations with maximum values $> 0.1$ are plotted.}
		\label{fig:lumo10}
	\end{figure}

\begin{figure}[h!]
\centering
		\includegraphics[width=0.8\linewidth]{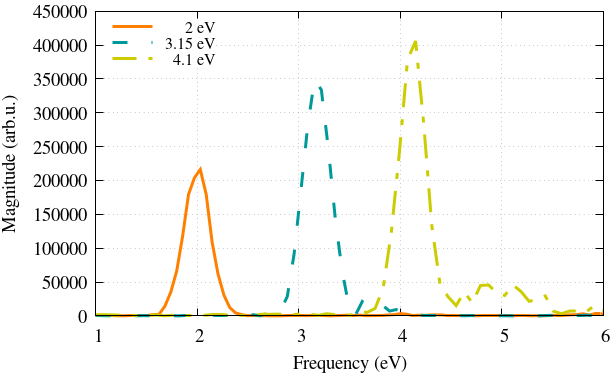}
		\caption{Fourier transform of the time-dependent dipole moment at field intensity $I_\mathrm{max} = 2 \times 10^{13}$ W/cm$^2$.}
		\label{fig:FT_dipole}
	\end{figure}

\begin{figure}
\centering
		\includegraphics[width=0.7\linewidth]{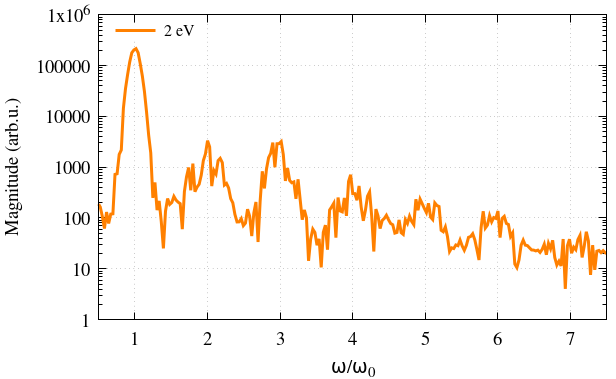}\\%
      \includegraphics[width=0.7\linewidth]{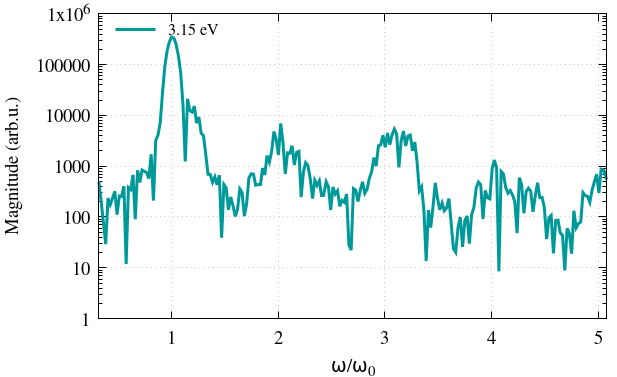}\\%
      \includegraphics[width=0.7\linewidth]{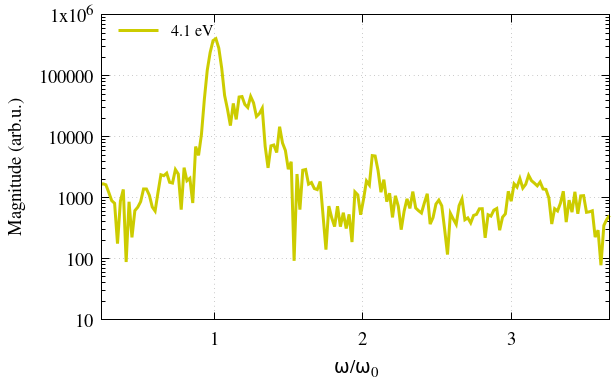}
		\caption{Fourier transform of the time-dependent dipole moment at field intensity $I_\mathrm{max} = 2\times 10^{13}$ W/cm$^2$ and frequencies (a) 2 eV, (b) 3.15 eV, (c) 4.1 eV. Higher harmonics are observed on a logarithmic scale.}
		\label{fig:FT_dipole_log}
	\end{figure}

	\begin{figure}[h!]
 \centering
		\includegraphics[width=0.5\linewidth]{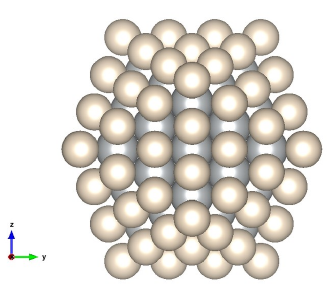}
		\caption{Atomic structure of the Ag$_{55}^{L1}$+H$_2$ surrounded by the layer of ghost atoms Ag$_\mathrm{g}$. Ag atoms in grey and Ag ghost atoms in white.}
		\label{fig:str-ghost}
	\end{figure}

 \begin{figure}
 \centering
		\includegraphics[width=0.8\linewidth]{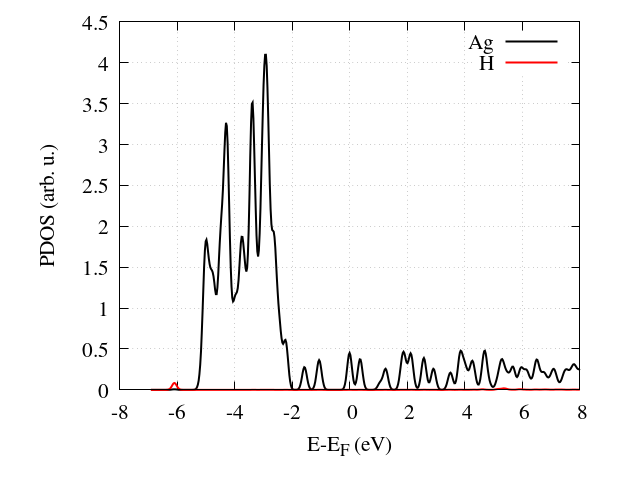}
		\caption{PDOS of the Ag$_{55}^{L1}+$H$_2$.}
		\label{fig:pdos-Ag-H}
	\end{figure}

  \begin{figure}
  \centering
		\includegraphics[width=0.8\linewidth]{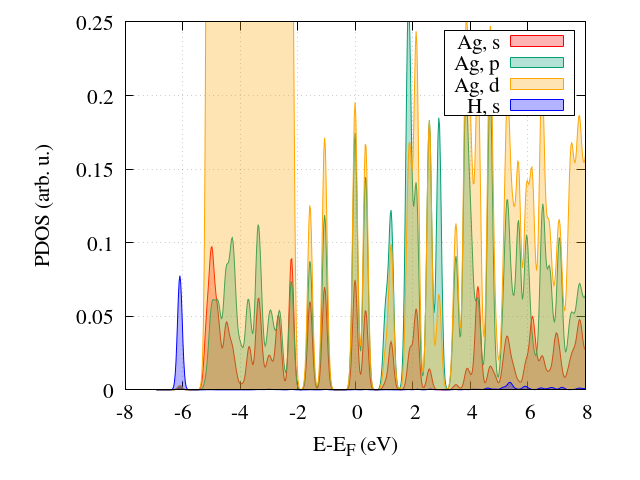}
		\caption{Angular-momentum resolved PDOS of the Ag$_{55}^{L1}+$H$_2$. The vertical axis scale is reduced to show the small features.}
		\label{fig:pdos-Ag-H}
	\end{figure}

 \begin{figure}
 \centering
		\includegraphics[width=0.8\linewidth]{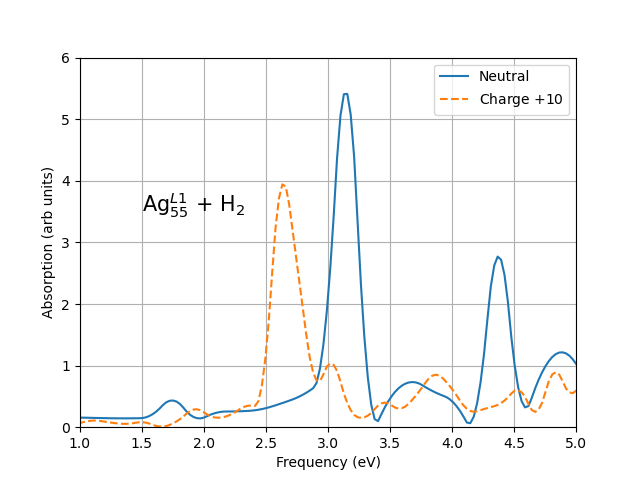}
		\caption{Absorption spectrum for neutral and charged system Ag$_{55}^{L1}+$H$_2$.}
		\label{fig:abs_spec_charged}
	\end{figure}